\documentstyle[preprint,prb,aps]{revtex}
\begin{document}
\input{psfig}
\title{Splitting the multiphase point}
\author{J. M. Yeomans}
\address{Theoretical Physics, Oxford University,
1 Keble Rd. Oxford OX1 3NP, UK}
\date{\today}
\maketitle
\begin{abstract}
Models with competing interactions, for example the ANNNI model, can
have special points at which the ground state is infinitely
degenerate, so-called multiphase points. Small perturbations can lift
this degeneracy and give rise to infinite sequences of long-period
phases. This paper compares the effect of three possible
perturbations, quantum fluctuations, thermal fluctuations and the
softening of the spins from their quantised positions.
\end{abstract}

\newpage

My aim in this paper is to summarise results showing that simple spin
models with short-range interactions can have surprisingly complex
phase diagrams, even at zero temperature. The models we shall consider
are three-dimensional, with ferromagnetic interactions in two of the
three dimensions. Along the third or axial direction there are
competing interactions which lead to the possibility of many
long-period structures becoming stable.

Arguably the paradigm of these systems is the axial
next-nearest-neighbour Ising or ANNNI model (Elliott 1961, Yeomans
1988, Selke 1988,1992) which has ferromagnetic
first-neighbour and antiferromagnetic second-neighbour interactions
along the axial direction. The ANNNI model Hamiltonian is
\begin{equation}
{\cal H}_A =
-J_0 \sum_{i \langle j j^{\prime} \rangle} \sigma _{i,j} \sigma _{i,j^{\prime}}
-J_1 \sum_{i, j } \sigma _{i,j} \sigma _{i+1,j}
+J_2 \sum_{i, j } \sigma _{i,j} \sigma _{i+2,j}.
\label{1}
\end{equation}
Here $J_1$ and $J_2$ are both positive, $\sigma _{i,j}=\pm 1$ is the
spin on site $(i,j)$, $i$ labels planes along the
axial direction, $j$ labels sites within a plane and $\langle j j^{\prime}
\rangle$ denotes a sum over nearest neighbours in a plane.

Defining $\kappa=J_2/J_1$ the ground state of the ANNNI model can
easily be seen by inspection to be ferromagnetic for $\kappa<1/2$ and an
antiphase structure consisting of two planes with spins $\sigma =1$
followed by two planes of spins with $\sigma =-1$ for $\kappa>1/2$.
$\kappa=1/2$ is a multiphase point
where the ground state is
infinitely degenerate with all possible combinations of
ferromagnetic and antiphase orderings having equal energy (Fisher and
Selke 1980,1981).

The existence of such a degeneracy leads one to suspect that small
perturbations could have a drastic effect on the phase diagram near
the multiphase point. Candidates are thermal fluctuations,
quantum fluctuations, or the softening of the spins
from the two discrete Ising values. We shall compare and contrast the
effect of the different perturbations. However, before considering
each of these
possibilities in turn, it is helpful to introduce the following
notation (Fisher and Selke 1980,1981).

Of the phases degenerate at the multiphase point  those that are
periodic can be labelled by using
$\langle n_1, n_2, \dots n_m \rangle$ to denote a state in
which the spins form domains (of parallel planes) whose widths
repeat periodically the sequence $n_1, n_2, \dots n_m$. For example
the antiphase state is labelled $\langle 2 \rangle$ and the phase in
which consecutive planes have the ordering $\ldots ++--++--- \ldots$
is $\langle 2223 \rangle$ or $\langle 2^3 3 \rangle$.
The term $p$-band will be used to describe $p$ consecutive planes of
up (down) spins terminated by down (up) planes.

\section*{Thermal fluctuations}

The phase diagram of the ANNNI model at finite temperatures near the
multiphase point was worked out some time ago
by Fisher and Selke (1980,1981) and refined by
Fisher and Szpilka (1987b). They found that a sequence of phases $\langle 2^k3
\rangle$, $k=0,1,2 \ldots k_{0}$ with $k_{0}\rightarrow \infty$ as the
temperature $T \rightarrow 0$ spring from the multiphase point.
Mixed phases $\langle 2^k32^{k+1}3 \rangle$ (and possibly more
complicated combinations of the basic sequences) are also stable at
any finite temperature for sufficiently large $k$. Fisher and Szpilka's
results are summarised in Figure 1.

The physics behind these results can be made transparent by
considering the spin system as an array of interacting domain
walls (Szpilka and Fisher 1986, Fisher and Szpilka 1987a).
There is some freedom
in the definition of a wall. For example
they  could be the boundaries between up and down bands but
here it is more convenient to consider the three-bands as walls within
a matrix of two-bands. At finite temperatures thermal wandering of the
walls leads to interactions between them which can differentially
stabilise the long-period phases.

The free energy of a given phase can be
written in terms of a
sequence of wall interaction free energies: $F_w$, the free
energy of an isolated wall; $V_2(n)$, the interaction
energy of two walls separated by $n$ sites; and generally
$V_k(n_1 , n_2 , \dots n_{k-1})$, the interaction energy of $k$
walls with successive separations $n_1$, $n_2$, ... $n_{k-1}$.
In terms of these quantities one may write the total free energy
of the system when there are walls at positions $m_i$ as
\begin{eqnarray}
F = && F_0 + n_w F_w + \sum_i V_2(m_{i+1}-m_i)
 +  \sum_i
V_3(m_{i+2}-m_{i+1},m_{i+1}-m_i)\nonumber \\ & + & \sum_i
V_4(m_{i+3}-m_{i+2},m_{i+2}-m_{i+1},m_{i+1}-m_i)
 +   \dots \ ,
\end{eqnarray}
where $F_0$ is the free energy with no walls present and $n_w$ is the
number of walls.

Successive approximations to the phase diagram follow from obtaining
the two-wall interactions, the three-wall interactions and so
forth. The stability of the phases in which the walls are equispaced
follows from a consideration of the pair interactions.
For a convex $V_2(n)$ all such phases are stable. Otherwise
the stable phases can be identified via a graphical construction due to
Fisher and Szpilka (1987a): if the extremal convex
envelope of $V_2(n)$ versus $n$ is drawn, the  points [$n,V_2(n)$] which
make up the envelope correspond to the stable phases in which the
walls are a distance $n$ apart which we shall denote $\{n\}$. (Note
that the notation $\{n\}$ only coincides with $\langle n \rangle$ for
the case where the walls have been identified as the edges of
ferromagnetic domains.)

The stability of the $\{n\}:\{n+1\}$
boundaries depends on the three-wall interactions. The
condition that the boundary correspond to a stable first-order transition is
that $F_n < 0$ (Fisher and Szpilka 1987a), where
\begin{equation}
\label{STAB}
F_n \equiv V_3(n,n) -2 V_3(n,n+1) + V_3(n+1,n+1) \ .
\end{equation}
If $F_n>0$, $\{n,n+1\}$ appears as a stable phase and there is the
possibility that four-wall interactions can stabilise $\{n,n,n+1\}$
[$\{n,n+1,n+1\}$] on the $\{n\}:\{n,n+1\}$ [$\{n+1\}:\{n,n+1\}$]
boundary and so on.

For the ANNNI model the wall--wall interactions were calculated by Fisher
and Szpilka (1987b) using a low temperature series expansion. The calculations
support the intuition that the graphs which give the leading order
contribution to $V_2(n)$ are chains of second-neighbour flipped spins
stretching between neighbouring walls.
Similar chains joining second-neighbour walls are responsible for
the three-wall interactions.
There are however subtleties which must be
addressed. For example disconnected graphs, which together span the
distance between the walls, must be taken into account and care must
be taken to subtract off the single wall energy and other background
contributions.

The leading order result is that $V_2(n)$ is
convex and $F_n<0$ leading to the conclusion that there is an infinite
phase sequence $\langle 2^k3\rangle$ and no mixed phases. However
for large $k \sim
\exp (4J_0/kT)$ Fisher and Szpilka (1987b) identified important correction
terms. These arise from diagrams when an extra spin is flipped within
the chain: although they carry an extra Boltzmann weight the number
of positions for the flipped spin increases with increasing $k$ and hence
these diagrams can eventually affect the phase diagram. They result in
a cut-off of the phase sequence and the appearance of mixed phases for
any finite temperature. Other corrections which arise from additional
flips outside the chain do not alter the structure of the phase
diagram.

\section*{Quantum fluctuations}

It is of interest to ask whether quantum fluctuations lead to a similar
splitting of the multiphase point of the ANNNI model. Obviously the
Hamiltonian (\ref{1}) is purely classical and cannot support quantum
fluctuations. Therefore we consider instead Heisenberg spins which
interact through the Hamiltonian
\begin{equation}
{\cal H}=
-\frac{J_0}{S^2} \sum_{i \langle j j^{\prime} \rangle}
{\bf S}_{i,j} \cdot {\bf S}_{i,j^{\prime}}
-\frac{J_1}{S^2} \sum_{i, j } {\bf S}_{i,j} \cdot {\bf S}_{i+1,j}
+\frac{J_2}{S^2} \sum_{i, j } {\bf S}_{i,j} \cdot {\bf S}_{i+2,j}
-\frac{D}{S^2} \sum_{i, j } ([ S_{i,j}^z]^2 -S^2)
\label{2}
\end{equation}
where ${\bf S}_{i,j}$ is a quantum spin of magnitude $S$. For
$D=\infty$ the Hamiltonian (\ref{2}) reduces to that of the ANNNI model.
For classical spins $S=\infty$ the ground state (and therefore
the multiphase point) is maintained for large $D$ and we shall work in
this limit.

To study quantum fluctuations Harris, Micheletti and Yeomans (1995a,b)
used the
Dyson-Maleev transformation
\begin{eqnarray}
S_i^z & = & \sigma_i ( S - a_i^+ a_i) \ , \nonumber \\
S_i^+ & = & \sqrt{2S} \left(
\delta_{\sigma_i,1} \left[ 1 - {a_i^+ a_i \over 2S} \right] a_i +
\delta_{\sigma_i,-1} a_i^+ \left[ 1-{a_i^+ a_i\over 2S} \right]
\right) \ , \nonumber \\
S_i^- & = & \sqrt{2S} \left( \delta_{\sigma_i,1} a_i^+
+ \delta_{\sigma_i,-1} a_i \right) \ ,
\end{eqnarray}
where $\delta_{a,b}$ is unity if $a=b$ and is zero otherwise
and $a_i^+$ ($a_i$) creates (destroys) a spin excitation at
site $i$, to  transform the Hamiltonian (\ref{2})
into the bosonic form
\begin{equation}
\label{HAM}
{\cal H} ( \{ \sigma_i \} ) = {\cal H}_A + {\cal H}_0
+ V_{||} + V_{\not{\parallel}} + {\cal O} (S^{-2}) \ ,
\end{equation}
where
\begin{equation}
{\cal H}_0 = \sum_{i,j} \Biggl[  2(D+2J_0) + J_1 \sigma_{i,j} ( \sigma_{i-1,j}
+
\sigma_{i+1,j} )
 - J_2 \sigma_{i,j} ( \sigma_{i-2,j} + \sigma_{i+2,j} ) \Biggr]
S^{-1} a_{i,j}^+ a_{i,j}
\end{equation}
with $\tilde{D}=D+2 J_{0}$
and $V_{||}$ ($V_{\not{\parallel}}$) is the interaction between spins
which are parallel (antiparallel)
\begin{eqnarray}
V_{||} = && {1 \over S} \sum_{i,j} \Biggl[ - J_1
X(i,i+1;j)
(a_{i,j}^+ a_{i+1,j} + a_{i+1,j}^+ a_{i,j} )
 + J_2
X(i,i+2;j)
(a_{i,j}^+ a_{i+2,j} + a_{i+2,j}^+ a_{i,j} ) \Biggr] \ ,
\end{eqnarray}
\begin{eqnarray}
V_{\not{\parallel}} = && {1 \over S} \sum_{i,j} \Biggl[ - J_1
Y(i,i+1;j)
(a_{i,j}^+ a_{i+1,j}^+ + a_{i+1,j} a_{i,j} )
+ J_2
Y(i,i+2;j)
(a_{i,j}^+ a_{i+2,j}^+ + a_{i+2,j} a_{i,j} ) \Biggr] \ ,
\end{eqnarray}
where $X(i,i';j)$ [$Y(i,i';j$)] is unity if spins $(i,j)$ and
$(i',j)$ are parallel [antiparallel] and is zero otherwise.

Just as for the case of thermal fluctuations it is helpful to think in
terms of an array of walls whose interactions now result from quantum
fluctuations. However, in contrast to finite temperatures, it is now
most convenient to define a wall as a boundary between bands.
The phases with equispaced walls which may be stabilised by the
two-wall interaction are then $\langle k \rangle$.

The form of the Hamiltonian (\ref{HAM}) allows us to immediately identify
the excitations responsible for mediating the wall--wall interactions by
noting that fluctuations out of the classical ground
state (the boson vacuum) can only be excited in pairs at the walls
by the perturbation
$V_{\not{\parallel}}$. $V_{||}$ is then able to propagate any
excitation within the ferromagnetic domains.

Hence the lowest order contribution to the two-wall interactions will
correspond to a pair of excitations at one wall, one of which moves to
the neighbouring wall {\em and back} and is then destroyed at the first
wall in tandem with its original partner. Similarly the graphs
responsible for the three-wall interactions correspond to a pair of
excitations at the centre wall each of which moves to its neighbouring
wall and back before being annihilated.

The contributions of these diagrams follow from standard
non-degenerate perturbation theory. The small parameter is
$J/(D+2J_0)$ ($J \equiv J_1$ or $J_2$). The result is that to leading
order the two-wall interaction is (Harris, Micheletti and Yeomans 1995b)
\begin{eqnarray}
V_2(n)&=&\frac{4J_2^nS^{-1}}{[4(D+2J_0)]^{n-1}}, \ \ \
n\  {\rm odd}, \nonumber \\
V_2(n)&=&\frac{J_2^{n-1}S^{-1}}{[4(D+2J_0)]^{n}} (n^2 J_{1}^{2}
-4 J_1J_2+8J_2^2)
, \ \ \ n \ {\rm even}.
\end{eqnarray}
Hence to leading order
$V_2(n)$ is a convex function of $n$
and all the phases $\langle k \rangle$ are stable.
$F(n)$ defined by equation
(\ref{STAB}) can be shown to be negative and the $\langle k \rangle :\langle
k+1 \rangle$
phase boundaries are first order.
Note that although this is qualitatively the same as the ANNNI
behaviour, the quantitative nature of the phase sequence is entirely
different.

Just as for the ANNNI model correction terms may be important for
large
$k \sim [(D+2J_0)/J]^{1/2}$. Firstly $V_2(n)$ will
suffer from strong even--odd
oscillations and therefore transitions $\langle k \rangle \rightarrow
\langle k+2 \rangle$ will appear. Secondly perturbations which follow
more complicated paths, although individually less important, may become
dominant because of their greater statistical weights. A calculation
attempting to take this into account indicates however that, unlike
the ANNNI model, the phase sequence does not terminate at a finite
value of $k$ (Harris, Micheletti and Yeomans 1995b).

This difference can be understood as follows.
In the present model in order for an excitation to sense the
presence of a second wall, it has to travel from one wall to
the other wall {\em and} return. Thus the interaction in the quantum
case is proportional to the square of an oscillatory Green's function, whereas
in the ANNNI model the analogous
function appears linearly.

\section*{Spin softening}

A third mechanism which might split the degeneracy at a multiphase
point is the softening of the spins themselves (that is a non-infinite
spin anisotropy $D$ in the Hamiltonian (\ref{2})). There is no splitting
for the ANNNI model because there is a finite energy barrier preventing
spins from moving continuously from their positions at
$D=\infty$. However, for some value of $D$ long-period commensurate or
incommensurate phases must be stabilised as, for $D=0$, the ground
state of the Hamiltonian (\ref{2}) is either ferromagnetic or
incommensurate with a wavevector that varies continuously with
$\kappa$.
Preliminary numerical results confirm that this is indeed the
case (Micheletti, 1995).

However a more complete description of the same physics exists
for a similar model
(Seno, Yeomans, Harbord, Ko 1994)
and it is this we prefer to consider here.
This is the classical X-Y model with first- and second-neighbour
competing interactions and a $p$-fold spin anisotropy $D$.
Each classical XY spin vector lies in a plane perpendicular to the
axial direction
and has unit magnitude. The Hamiltonian is
\begin{equation}
{\cal H}_{XY}= - J_{0}
\sum_{i \langle j j^{\prime} \rangle }{\bf s}_{i,j}\cdot{\bf s}_{i,j^{\prime}}-
J_{1} \sum_{i,j} {\bf s}_{i,j}\cdot{\bf s}_{i+1,j}+
J_{2} \sum_{i,j} {\bf s}_{i,j}\cdot{\bf s}_{i+2,j}
+ D \sum_{i,j}\bigg(1-\cos(6\theta_{i,j})\bigg)
\label{clock}
\end{equation}
where $\theta_{i,j}$ is the angle between the spin at site $(i,j)$ and
a given axis.
This model is relevant to an understanding of the ferrimagnetic
ordering of  rare-earths such as holmium where the spins are confined to the
basal plane and subjected to a hexagonal spin anisotropy (Jensen and
Mackintosh 1991).

The ground state of the Hamiltonian (\ref{clock}) is well understood in
the two limits $D=0$ and $D=\infty$. For $D=0$ the ground state is
ferromagnetic for $\kappa < {1 \over 4}$. For $\kappa > {1 \over 4}$
it exhibits helical order with a wavevector which
is, in general, incommensurate with the underlying lattice. The
magnitude of the wavevector is determined by the exchange energies
through the relation $\cos{q}=(4 \kappa)^{-1}$.

For $D=\infty$, however, the spin angles $\theta_{i,j}$ are constrained to
take  one of the discrete set of  values $\pi k_{i,j}/3$,
where $k_{i,j}=0,1,2,3,4,5$ will be used to label the different spin states.
The Hamiltonian (\ref{clock}) then reduces to the $6$-state clock model with
competing interactions.
The ground state now has a very different character: only a few short-period
commensurate phases are stable as $\kappa$ is varied. For $\kappa < 1/3$
the ground state is ferromagnetic. For ${ 1 \over 3} < \kappa < 1$
the order along the axial direction is
helical with a sequence $k_{i} \equiv k_{i,j}=\ldots 01234501 \ldots $, with
spins in adjacent planes differing by an angle $(\pi / 3 )$. For
$ \kappa > 1$ there are two degenerate states at zero temperature
$\ldots 01340134 \ldots$
and $\ldots 00330033 \ldots$.
Our aim is to describe the ground state of the Hamiltonian
(\ref{clock}) as a function of $D$ and in particular the crossover
between the two very different
types of ordering at $D=0$ and $D=\infty$.

For $D=\infty$ the ferromagnetic phase
$\langle \infty \rangle$
and helical phase $\langle 1 \rangle$ coexist
for $\kappa=1/3$. However, this is not a multiphase point and
there is a first-order transition between the phases for
large $D$. Decreasing $D$ neither $\langle \infty \rangle$
nor $\langle 1 \rangle$ change their energy as the spins
remain along an easy axis.
Consequently the transition remains at $\kappa=1/3$.

However, it is also necessary to consider two sets of phases
which in the limit
$D=\infty$ are very close in energy to the ferromagnetic
and helical phases but
which  lower their energy as $D$ decreases.
At ($\kappa=1/4$,  $D=\infty$) all structures obtained by
combining $1$- and $2$-bands
are degenerate.
At ($\kappa=1/2$, $D=\infty$) all phases comprising $m \geq 2$
bands are degenerate. As $D$  decreases sequences
of periodic phases spring from the multiphase points.
For large $D$, however, these are metastable because the phases
$\langle\infty\rangle$ and $\langle 1\rangle $ have
lower energies. However, unlike
$\langle \infty\rangle $ and $\langle 1\rangle $ they
can decrease their energy by a canting
of the spins. For example in the phase
$\langle 21\rangle $ the two parallel spins
move apart as $D$ is decreased reaching, for $D=0$, the
uniform arrangement with $q=2 \pi  / 9$. Therefore
we have to consider the possibility of their appearing
as stable phases for small $D$.

Following the energies of the hidden phase sequences
numerically it is apparent that this is indeed
the case (Seno, Yeomans, Harbord, Ko 1994).
The results are shown in Figure 2a and are
enlarged in Figure 2b.
The phase diagram can be built up inductively with each phase being
constructed from its neighbours (eg $\langle 12 \rangle$
 $+$ $\langle 12^2 \rangle$ $\rightarrow$ $\langle 1212^2 \rangle$)
as is usual for models of this type.
Within numerical limitations
 all the expected states appear.
Higher order phases are expected to occupy extremely small regions of
the phase diagram and cannot be resolved numerically.

A different behaviour is seen near $\kappa=1$ which is a multiphase
point (see Figure 2).
Here all states for which $|k_{i+1}-k_{i}| = 1 $ or $2$,
with the proviso
that two neighbouring jumps of $2$,
$|k_{i+2}-k_{i+1}|=|k_{i+1}-k_{i}|=2$, are
forbidden, are stable.
We define a wall as lying between sites $i$ and $i+1$
if $|k_{i+1}-k_{i}|=2$. The notation used above can still be employed
to describe the stable states but we use square brackets to indicate
that a band now contains helically coupled spins.
 For example $ \ldots 01\bigg|345\bigg|12\bigg|450\ldots$,
where walls are denoted by vertical lines,
 will be
labelled  $[ 23 ]$.

A $1/D$ expansion (Seno and Yeomans 1994) shows that all phases which only
contain  bands
of length $\geq 3$ and which obey the branching rules,
spring from the multiphase point at $\kappa=1$.
To within the accuracy of the numerical calculation all phases
containing $2-$ and $3$-bands then appear between $[2]$ and $[
3]$ as $D$ is decreased. It was possible to check for the existence of
 phases with periods of up to $100$ lattice spacings.

The solid  phase boundaries shown in Figure 2 follow the numerical results.
  As $D \to 0$ we also  show by dotted lines the expected behaviour;
that the phase widths
decrease and a given phase touches the $D=0$ axis at a single
 point corresponding to the appropriate value of $q$.
Phases arising from the multiphase points at
$\kappa=1/4$, $\kappa=1/2$, and $\kappa=1$ touch the $D=0$ axis for
ranges of $\kappa$ from $1/2 \sqrt{3} \to 1/2$, $1/4 \to 1/2 \sqrt{3} $,
and $1/2 \to \infty$ respectively.
It is not possible to follow the low anisotropy behaviour numerically because
an infinite number of phases would have to be considered.

An important question is whether incommensurate phases persist in the
phase diagram for non-zero spin anisotropy. In the continuum limit the
Hamiltonian (\ref{clock}) can be mapped onto the Frenkel-Kontorova
model. Thus
it is expected on the basis of previous work that,
for small $D$, the
devil's staircase is incomplete with incommensurate phases appearing
between the commensurate ones (Bak 1982).

\vspace{1cm}

In conclusion I should like to emphasise the immense richness of
behaviour seen in these simple spin systems. The degeneracy of the
multiphase point allows any perturbation to have a strong effect on
the phase diagram. Although there are many qualitative similarities
between the effect of different perturbations the quantitative
results are strongly
dependent on the details of the perturbation and  the underlying
model.

Finally I should point out that the models described here are of more
than just theoretical interest. For example they have been used to
model the ferrimagnetic order of the rare earths, mineral polytypism
and antiphase domain ordering in binary alloys (Jensen and Mackintosh
1991, Loiseau, Van Tendeloo, Portier and Ducastelle 1985, Cheng, Heine
and Jones 1990).

\vspace{1cm}

ACKNOWLEDGEMENTS:
I should like to thank
M.E. Fisher, A.B. Harris, C. Micheletti, and F. Seno, with whom I
have enjoyed working on problems in this field.
I acknowledge the support of an EPSRC Advanced Fellowship.

\newpage
\section*{References}
{}~\\
P. Bak, Commensurate phases, incommensurate phases and the devil's
staircase.
Rep. Prog. Phys. {\bf 45} 587 (1982).\\
{}~\\
C. Cheng, V. Heine and I.L. Jones, SiC polytypes as equilibrium
structures. J. Phys. Cond. Mat. {\bf 2} 5097 (1990).\\
{}~\\
R.J. Elliott,
Phenomenological discussion of magnetic ordering in the heavy
rare-earth metals.
Phys. Rev. {\bf 124} 346 (1961).\\
{}~\\
M.E. Fisher and W. Selke,
Infinitely many commensurate phases in a simple Ising model.
Phys. Rev. Lett. {\bf 44} 1502 (1980).\\
{}~\\
M.E. Fisher and W. Selke, Low temperature analysis of the
axial next-nearest neighbour Ising model near its multiphase
point.  Philos. Trans. R. Soc. (London), A{\bf 302} 1 (1981).\\
{}~\\
M.E. Fisher and A.M. Szpilka, Domain wall interactions I General
features and phase diagrams for spatially modulated phases.
Phys. Rev. B {\bf 36} 644 (1987a).\\
{}~\\
M.E. Fisher and A.M. Szpilka, Domain-wall interactions II
High-order phases in the axial next-nearest-neighbour Ising
Model.  Phys. Rev. B {\bf 36} 5343 (1987b).\\
{}~\\
A.B. Harris, C. Micheletti and J.M. Yeomans, Quantum fluctuations in
the axial next-nearest neighbour Ising
model, Phys. Rev. Lett. {\bf 74} 3045 (1995a).\\
{}~\\
A.B. Harris, C. Micheletti and J.M. Yeomans,
Lifting of multiphase degeneracy by quantum fluctuations.
Submitted to Phys. Rev. E. (1995b)\\
{}~\\
J. Jensen and A.R. Mackintosh  {\em Rare Earth Magnetism and
Excitations.}
(Oxford University Press, Oxford 1991).\\
{}~\\
A. Loiseau, G. Van Tendeloo, R. Portier and F. Ducastelle,
Long-period structures in Ti$_{1+x}$Al$_{3-x}$ alloys -- experimental
evidence of a devil's staircase.
J. Phys. (Paris) {\bf 46} 595 (1985).\\
{}~\\
C. Micheletti.  Private communication (1995)\\
{}~\\
W. Selke, The ANNNI model -- theoretical analysis and experimental
applications. Phys. Rep. {\bf 170} 213 (1988).\\
{}~\\
W. Selke in {\em Phase transitions and Critical phenomena}
{\bf 15} eds. C. Domb and J.L. Lebowitz (New York: Academic, 1992).\\
{}~\\
F. Seno, J.M. Yeomans, R. Harbord and D.Y. K. Ko, Ground state
of a model with competing interactions and spin
anisotropy.  Phys. Rev. B {\bf 49} 6412 (1994).\\
{}~\\
F. Seno and J.M. Yeomans, Spin softening
in models with competing interactions: a new high anisotropy
expansion to all orders.  Phys. Rev. B {\bf 50} 10385 (1994).\\
{}~\\
A.M. Szpilka and M.E. Fisher, Domain-wall interactions and
spatially modulated phases.  Phys. Rev. Lett {\bf 57} 1044 (1986).\\
{}~\\
J.M. Yeomans, The theory and application of axial Ising models.
Solid State Physics, {\bf 41} 151 (1988).\\

\section*{Figure Captions}
{}~\\
{}~\\
\begin{itemize}

\item[Figure 1:] Schematic phase diagram of the ANNNI model at
low temperatures.  The mixed phases $\langle 2^k 32^{k+1} 3 \rangle$
may be unstable to the appearance of higher-order mixed
phases.  The phase widths decrease exponentially with $k$ and,
for clarity, the widths of the higher-order phases have
been exaggerated. (After Fisher and Szpilka 1987 b.)

\item[Figure 2(a):] Ground state phase diagram of the XY
model with competing axial interactions and 6-fold spin anisotropy
$D$.  Bold lines depict the numerical results; the dotted
boundaries show the expected behaviour of the phase boundaries
as $D \to 0$.

\item[Figure 2(b):]  An enlargement of Figure 2(a) for
$1/4 < J_2/J_1 < 1/2$ and small $D$.
(After Seno, Yeomans, Harbord and Ko 1994.)

\end{itemize}
\end{document}